\documentclass[]{aa}  
\usepackage{graphicx,subfigure,natbib,url,amsmath,mathtools,lscape}
\usepackage[varg]{txfonts}
\def\modz{\left| z \right|}
\begin{document}

\title{A homogeneous sample of 34\,000 M7$-$M9.5 dwarfs brighter than
$J=17.5$ with accurate spectral types\thanks{The catalogue (Table 1) is only available at the CDS via anonymous ftp to cdsarc.u-strasbg.fr (130.79.128.5) or via http://cdsarc.u-strasbg.fr/viz-bin/qcat?J/A+A/}}

\titlerunning{A homogeneous sample of 34\,000 M7$-$M9.5 dwarfs brighter than
$J=17.5$}


\author{S. Ahmed,  \and S.J. Warren}
\authorrunning{S. Ahmed \& S.J. Warren}

\institute{Astrophysics Group, Imperial College London, Blackett
  Laboratory, Prince Consort Road, London SW7 2AZ, UK \label{inst1}}

\date{Received <date> / Accepted <data>}

\abstract{The space density of late M dwarfs, subtypes M7 to M9.5, is
  not well determined. We applied the photo-type method to {\em iz}
  photometry from the Sloan Digital Sky Survey and {\em YJHK}
  photometry from UKIRT Infrared Deep Sky Survey, over an effective
  area of 3070\,deg$^2$, to produce a new, bright
  $J\mathrm{(Vega)}<17.5$, homogeneous sample of 33\,665 M7 to M9.5
  dwarfs. The typical S/N of each source summed over the six bands is
  $>100$. Classifications are provided to the nearest half spectral
  subtype. Through a comparison with the classifications in the BOSS
  Ultracool Dwarfs (BUD) spectroscopic sample, the
  typing is shown to be accurately calibrated to the BUD
  classifications and the precision is better than 0.5 subtypes rms;
  i.e. the photo-type classifications are as precise as good spectroscopic
  classifications. Sources with large $\chi^2>20$ include several
  catalogued late-type subdwarfs. The new sample of late M dwarfs is
  highly complete, but there is a bias in the classification of rare
  peculiar blue or red objects. For example, L subdwarfs are
  misclassified towards earlier types by approximately two spectral
  subtypes. We estimate that this bias affects only $\sim1\%$ of the
  sources. Therefore the sample is well suited to measure the
  luminosity function and investigate the softening towards the
  Galactic plane of the exponential variation of density with height.}


\maketitle

%

\section{Introduction}

A detailed census and study of the coolest stellar objects, i.e. late
M dwarfs and cooler, first became possible with the implementation of
wide-field surveys at wavelengths $0.8-2.4\mu$m, especially the Sloan
Digital Sky Survey (SDSS; \citealt{york00}) and the Two Micron All Sky
Survey (2MASS; \citealt{2MASS}). The highlight in this area was the
discovery and characterisation of the L and T dwarf populations
\citep{kirkpatrick99,martin99,strauss99,geballe02,burgasser02,burgasser06b,hawley02,schmidt10}. At
the same time new insights into the properties of M dwarfs have been
obtained, including the quantification of their activity as a function
of spectral type and age \citep{west08,west11,schmidt15}, as well as
measurement of the luminosity function
\citep[LF;][]{cruz07,covey08,bochanski10}. From the LF, using a
relation between luminosity and mass \citep[e.g.][]{delfosse2000}, the
stellar initial mass function (IMF) may be derived. The LF of M dwarfs
is important because the characteristic mass of a log-normal fit to
the IMF lies within this spectral range
\citep{chabrier03,bochanski10}.

The study by \citet{bochanski10} is the most complete analysis of the
M dwarf LF. They determined new photometric parallax relations
(absolute magnitude as a function of colour) and applied these values
to a sample of $\sim15\times10^6$ M dwarfs from 8,400\,deg$^2$ of SDSS
Data Release 7 to derive the LF over the absolute magnitude interval
$7<M_{r(AB)}<16$. Their LF varies smoothly over the interval
$7<M_{r(AB)}<14$, corresponding to spectral types M0 to M5, but
displays significant fluctuations between bins for absolute magnitudes
$M_{r(AB)}>14$. This is partly owing to the relatively small numbers
at these absolute magnitudes, which is a consequence of their use of
the $r$ band for the sample magnitude limit, $r(AB)=22$. Because late
M dwarfs, $>$M5, are so red, imposing a cut in $r$ leads to a rapid
reduction in limiting distance, and thus sample size, towards later
spectral types. The space densities for the latest M dwarfs are
additionally uncertain because the photometric parallax relation is
less well calibrated in this region. The measured IMF is further
impacted because the colour correction from the $r$ band to the $J$
band, which is the passband of the \citet{delfosse2000}
mass-luminosity relation, is large for late M dwarfs and not precisely
determined \citep{hawley02}.

An independent estimate of the M dwarf LF for spectral types M7 to M9
was made by \citet{cruz07}, using a sample of 53 stars within 20pc of
the Sun. The objects were identified using 2MASS $JHK_s$ photometry by
application of a colour cut $J-K_s>1.0$. \citet{cruz07} estimated that
$79\%$ of M7 dwarfs are redder than this colour limit and all M8 and
M9 dwarfs. Unfortunately this sample is also problematic. Recent
analysis of the 2MASS colours of M dwarfs by \citet{schmidt15}
provided median colours $J-K_s=0.96, 1.03$ for M7, M8. These results
suggest that only $\sim50\%$ of M7 and M8 dwarfs satisfy the above
colour cut, meaning their space densities have been substantially
underestimated.

These questions motivate the compilation of a new sample of late M
dwarfs, hereafter M7$-$M9.5, in order to obtain an improved
measurement of the LF. We now briefly consider the issues involved in
making an accurate measurement of the LF. Various studies of the LF
may be distinguished by whether distances are measured by
trigonometric, spectroscopic (absolute magnitude as a function of
spectral type), or photometric parallax. The precision of the
distances decreases along this sequence, but this is usually
compensated by the increase in sample size. Trigonometric parallaxes
can only be measured for the nearest, brightest sources, and similarly
close binaries can only be resolved for the nearest sources. In
comparison, for the largest samples that use photometric parallax,
the distances are less precise and a statistical correction for
unresolved binaries must be made. The variation in space density over
the volume surveyed, because of the structure of the Milky Way, further
complicates the measurement of the LF. These
factors mean that comparison between local \citep[such
  as][]{cruz07} and distant \citep[such as][]{bochanski10}
determinations of the LF is usually not straightforward. For
reference, about one quarter of the \citet{cruz07} sources benefit
from a trigonometric parallax, while the remainder have spectroscopic
parallaxes.

In this paper we present a new large homogeneous sample of 34\,000
M7$-$M9.5 dwarfs. We use ``homogeneous'' to mean that the sample has
high completeness and for which the incompleteness is accurately
quantified. The new sample exploits the deeper UKIRT Infrared Deep Sky
Survey (UKIDSS) photometry \citep{UKIDSS}, compared to
2MASS, and combines some of the respective advantages of the two
previous studies, in that it is not only large but also benefits from precise distances. The
sample approximates a complete spectroscopic sample: we use the
phototype method of \citet{skrzypek15,skrzypek16}, combining SDSS $iz$
and UKIDSS $YJHK$ photometry, to measure accurate spectral types. The
spectroscopic parallax relation of \citet{dupuy12} then provides
distances precise to $15\%$. All the M dwarfs in the new sample lie
within 235\,pc of the Sun. In a companion paper (Warren and Saad, in
prep.) we will analyse the sample to measure the space density as a
function of spectral type, which is equivalent to the LF.

The layout of the remainder of the paper is as follows. We describe
the sample selection in \S\ref{selection}. The sample and its main
characteristics are presented in \S\ref{sample}. In \S\ref{accuracy}
we quantify the precision of the spectral typing. We summarise in
\S\ref{summary}.

\section{Selection}
\label{selection}

\subsection{Photo-type method}

The photo-type method \citep{skrzypek15} uses multiband
photometry to measure spectral types. The wide wavelength coverage can
compensate for the very low wavelength resolution of broadband
photometry, and with high signal-to-noise ratio data photo-type
classifications of late-type dwarfs are competitive with spectroscopic
classifications in precision of spectral typing. \citet{skrzypek15}
developed the method using SDSS+UKIDSS+ALLWISE $izYJHKW1W2$ 8-band
photometry for the discovery of L and T dwarfs in wide-field survey
data. Using a set of stars and brown dwarfs classified by standard
spectroscopic methods, polynomial relations between colour and
spectral type were determined for the seven colours\footnote{Revised
  template colours for M7-M9 dwarfs were published by
  \citet{skrzypek16}}. A source is then classified by fitting the
spectral energy distribution (SED) against the set of templates for
each MLT spectral subtype, as well as a range of templates for
quasars and other possible contaminants, selecting the min$-\chi^2$
fit as the classification. The $\chi^2$ of the best fit is useful for
identifying sources with spurious photometry and interesting
peculiar objects.

In \citet{skrzypek16} the method was applied to the classification of point sources in the magnitude range $13.0<J<17.5$, with colours $Y-J>0.8$, resulting in a sample of 1281 L and 80 T dwarfs, from an effective area of 3070\,deg$^2$. The matching radius criteria (within UKIDSS, and in matching to SDSS and WISE) ensures that incompleteness due to proper motion is negligible. The effective area calculation accounts for sources lost due to unreliable photometry in any of the bands from a variety of causes. The relative depths in the different bands, the matching criteria, and the $J$ magnitude range mean that incompleteness due to the requirement that a source is detected in all of the $YJHK$ bands is negligible. The result is that the sample is essentially complete for all spectral types L0 to T8, other than a small incompleteness that is estimated at $3\%$ because peculiar blue L dwarfs are classified M and a related overcompleteness because peculiar red M dwarfs are classified L. 

\subsection{Selection of M7 to M9.5 dwarfs}

The goal of the current paper is to extend the survey of
\citet{skrzypek16} to earlier spectral types M7 to M9.5. The methods
used are almost identical to those previously used. Small
modifications to this approach mostly result from the difficulty to
scrutinise all the images of all the objects in the much larger
sample of over 30\,000 late M dwarfs compared to 1361 L and T
dwarfs. In this work we provide only a brief outline of the methods, deferring
to the earlier papers for details, but we explain any
differences in depth. Because we used almost identical procedures for the
sample of late M dwarfs, we assume that the effective area calculated
for the LT sample, 3070\,deg$^2$, also applies to the M dwarfs. We
also assume that contamination of the sample by giants is negligibly
small based on the arguments presented by \citet{ferguson17}.

The photometric bands used in this study are the $i$ and $z$ bands in
SDSS and the $YJHK$ bands in UKIDSS. All the magnitudes and colours
quoted in this paper are Vega based, unless explicitly labelled as AB,
by for example $r(AB)$. The $YJHK$ survey data are calibrated to Vega,
while SDSS is calibrated on the AB system. We applied the offsets
tabulated in \citet{hewett06} to convert the SDSS $iz$ AB magnitudes
to Vega.

Starting with a set of point sources in the UKIDSS Large Area Survey
data release 10, detected in all four bands, $YJHK$, and in the range
$13.0<J<17.5$, we matched to SDSS DR9 using a $10\arcsec$ match
radius and selected the point spread function (psf) photometric measurements. We note that a M7
dwarf has colours $i-z=1.36$, $Y-J=0.68$ \citep{skrzypek16}, and so we apply
colour cuts $i-z>1.0$, $Y-J>0.4$ to reduce the total number of
sources, while ensuring all sources M7 and later are retained (this
assumption is checked later).  To allow for sources with significant
proper motion, we match to the nearest SDSS source that does not have
a closer match to a different UKIDSS source. Considering the $izYJHK$
colours of M7 to M9.5 dwarfs \citep{skrzypek16}, and the limiting
depths in the different bands, sources with $13.0<J<17.5$ are
easily detected in all the other bands. A small number of the faintest
M7 to M9.5 dwarfs do not have photometry in the ALLWISE W1 and W2
bands because the sources are undetected in both bands; a source only needs to be
above $5\sigma$ in one band to be measured in both bands. In any
case the colours $K-W1$ and $W1-W2$ vary very little over this
spectral range and therefore add essentially no useful information to
the classification of late M dwarfs. Recalling that some $7\%$ of
sources are blended in the WISE images \citep{skrzypek16}, which would
all have to be identified by eye, we decided not to match to ALLWISE
and  limited our analysis to the $izYJHK$ photometry.

This initial sample contained 404\,496 sources. We used template colours for L0 to T8 from \citet{skrzypek15} and for M7 to M9 from \citet{skrzypek16}.  We used newly determined colours for M0 to M6 from Barnett et al. (in prep). We classified all sources to the nearest half spectral subtype by interpolating the colours. The final sample of M7 to M9.5 dwarfs, after quality control, contains 33\,665 sources.

A distinct difficulty with starting with a near-infrared catalogue and
matching to an optical catalogue to identify genuinely cool stellar
objects is that hotter objects with erroneously faint photometry in
the SDSS $iz$ bands will be selected as candidates.  With the L and T
sample of \citet{skrzypek15} it was possible to identify these by
scrutinising every candidate, but this is not possible with the much
larger late M dwarf sample. This leads to a detailed consideration of
the use of the data quality flags provided for all SDSS sources in
each band. We first eliminated sources if either of the flags 
 \verb PSF_FLUX_INTERP \ or \verb BAD_COUNTS_ERROR \ were set in either $i$
 or $z$. We also removed any candidates that were close to a bright
 star $J<11$ in the 2MASS catalogue, using the criterion
 $\theta\arcsec<108-8J$. This relation was derived by first selecting
 sources in the catalogue with large $\chi^2$, and plotting the
 angular separation to 2MASS sources brighter than $J=11$, against $J$
 of the 2MASS source. A clearly defined locus presented itself,
 demonstrating that the photometry of sources within this locus is not
 reliable.

After this filtering we started to scrutinise sources with poor SED
fits. We identified many sources with incorrect photometry, i.e. a
total of $\sim1100$ by the end or $\sim3\%$ of the initial late M
dwarf sample. Nearly all of these were recorded in SDSS as very faint
in $i$, whereas they are in fact clearly visible in the SDSS images,
meaning that the SDSS $i$ photometry is wrong. In addition, in a
number of cases the UKIDSS source was matched to the wrong SDSS source
because the star, although visible in the SDSS images, was not listed
in the SDSS catalogue. In such cases, typically the UKIDSS source
became matched to a faint nearby galaxy in SDSS. Since by the nature
of the selection we are picking up a significant proportion of all SDSS
sources that are either completely missed or measured spuriously
faint, this is a very minor problem for SDSS sources in general; but
it is an important problem in selecting cool M dwarfs, starting with a
UKIDSS source list. We found that the
other recommended flags for identifying doubtful photometry
\citep[see, { e.g.,}][]{covey08} were ineffective at identifying most
of these spurious sources. The majority of the bad sources are
products of deblending from a brighter neighbouring source, but
without any of those flags being set. It seems that very occasionally
the deblending algorithm produces incorrect results.

To deal with objects with bad photometry from whatever cause, we
analysed further all sources with $\chi^2>10$. We first ranked the
sources on $\chi^2$, plotted their SEDs, and identified sources
where the photometry did not agree with our visual inspection of the
SDSS and UKIDSS images. To deal with the particular problem of
incorrect $i$ photometry, we matched all sources with $\chi^2>10$ to
Pan-STARRS and compared the $i$ photometry between SDSS and
Pan-STARRS. Because the $i$ bandpasses of the two surveys are almost
identical \citep{tonry12} this is particularly useful. Spurious
sources were then identified as those for which the SDSS $i$
photometry was anomalous, both compared to Pan-STARRS and when
interpolating between $r$ and $z$ in SDSS. Following these procedures
a large percentage of all the sources that had $\chi^2>20$ were
eliminated, but by the time the threshold of $\chi^2=10$ was
approached almost all sources were classified as good. This implies
that the residual proportion of sources that have bad photometry in the
final catalogue is extremely small $\ll 1\%$.

One further issue to do with the SDSS flags is noteworthy. In the
final catalogue a number of sources are included that have the SDSS
flag \verb SATURATED \ set in either the $i$ or the $z$ band. In
producing a clean set of stellar sources from SDSS data it is common
to eliminate such sources, but we deliberately kept them in. Because
we set a bright limit $J>13$, late M dwarfs in our catalogue are not
saturated in SDSS. Given their colours, the brightest sources have
$i(AB)=16.0$, $z(AB)=14.8$, which is well below the saturation
limit. Sources in our catalogue that have the SDSS flag 
 \verb SATURATED \ set must have been deblended from a neighbouring
 saturated source and inherited the flag \citep[this point is noted
   by][]{covey08}. We have no reason to believe that the photometry is
 incorrect. Therefore the classifications should be reliable. We wish to
 retain these objects, because in some cases they are binary
 companions to the bright star from which they were deblended, and
 therefore could be valuable as benchmark systems.

We did not use ALLWISE photometry for the new sample, but it was used
for the LT sample of \citet{skrzypek16}. Therefore there is a slight
ambiguity in membership between the two samples for a handful of
sources at the M9.5/L0 boundary at the level of 0.5 subtypes. For
example we might classify some of their L0 sources as M9.5 (these
would then appear in both samples) and we might classify some sources
as L0 that they classified as M9.5 (these would then be absent from both
samples). For those few sources that cross the M9.5/L0 boundary
with/without ALLWISE, we resolved in favour of the classification that
used ALLWISE, ensuring that there is no inconcsistency between the sample in
this paper and the sample in \citet{skrzypek16}.

\section{Sample}
\label{sample}

The new sample is presented in Table \ref{mainsample}, sorted by right
ascension, listing in successive columns as follows: (1) the UKIDSS ICRS
coordinates, (2) the angular separation in arcsec to the SDSS match,
(3-14) the six-band $izYJHK$ photometry, (15) the photo-type
classification PhT (to the nearest half subtype), and (16) the
$\chi^2$ of the fit. Column (17) lists the quantity $E_8$, which is
the Bayestar17 \citep{green18} reddening computed for a distance
modulus of 8.0 (a distance of 400\,pc). This quantity is discussed
further below. It is an overestimate of the actual reddening for any
of the sources, as they all lie at smaller distances. The remaining
columns list: (18) the spectroscopic classification SpT if the object
is present in the BOSS Ultracool Dwarfs (BUD) sample of
\citet{schmidt15} (listed as 99 otherwise); (19) the distance in pc
estimated from PhT, using the relation between spectral type and
absolute magnitude provided by \citet{dupuy12} for the $J$ band; and (20,
21) the Galactic coordinates $l,b$. The total number of sources in the
sample is 33,665.

\begin{sidewaystable}\tiny
\caption{Sample of 33\,665 M7 to M9.5 dwarfs}
\begin{tabular}{cccccccccccccclccccrr}
\hline\hline
Name & sep. & {$i$} & {$\sigma_i$} & {$z$} & {$\sigma_z$} & {$Y$} & {$\sigma_Y$} & {$J$} & {$\sigma_J$} & {$H$} & {$\sigma_H$} & {$K$} & {$\sigma_K$} & {PhT} & {$\chi^2$} & {$E_8$} & {SpT} & dist. & \multicolumn{1}{c}{l} & \multicolumn{1}{c}{b}  \\
 & arcsec. & & & & & & & & & & & & & & & & & pc & \multicolumn{1}{c}{deg.} & \multicolumn{1}{c}{deg.} \\ \hline
ULAS\,J000000.07+151212.9 & 0.18 & 19.86 & 0.03 & 18.29 & 0.05 & 17.66 & 0.02 & 16.88 & 0.02 & 16.33 & 0.02 & 15.93 & 0.03 & M7.5 & 2.36 & 0.041 & 99 & 158.3 & 104.97 & $-45.87$ \\
ULAS\,J000001.27+104504.8 & 0.02 & 19.45 & 0.03 & 18.07 & 0.03 & 17.46 & 0.02 & 16.71 & 0.01 & 16.15 & 0.02 & 15.76 & 0.02 & M7.5 & 5.20 & 0.084 & 99 & 146.6 & 103.00 & $-50.12$ \\
ULAS\,J000002.81+081047.8 & 0.26 & 19.96 & 0.04 & 18.68 & 0.05 & 17.96 & 0.03 & 17.36 & 0.03 & 16.79 & 0.04 & 16.35 & 0.05 & M7 & 4.72 & 0.069 & 99 & 220.1 & 101.71 & $-52.56$ \\
\end{tabular}
\tablefoot{Only the first three lines of the table are provided. The full table is available at the CDS.} 
\label{mainsample}
\end{sidewaystable}

The sample is of high S/N and low reddening. To characterise the S/N of the data we list in Table \ref{photomerrors} the median uncertainty in each band and the $90\%$ quantile. The typical S/N is 30 in the optical bands and 50 in the near-infrared bands, providing a combined S/N over the six bands of over 100.
\citet{green18} have published three-dimensional maps of reddening $E(g-r)$ using Pan-STARRS data. The LAS areas are predominantly at high Galactic latitude. Using the absolute magnitudes provided by \citet{dupuy12}, the median distance of the objects in the sample is 159\,pc, and the objects all lie within 235\,pc of the Sun.  At these distances, at high-Galactic latitude, the number of stars available is too small for the algorithm of \citet{green18} to work well, therefore it is not possible to provide an accurate reddening for each source. Instead we list the estimated reddening $E_8$ for a distance modulus of 8.0, providing an upper limit to the actual reddening to highlight the few sources for which the reddening could affect the classification. The quantiles of $E_8$ are also provided in  Table \ref{photomerrors}. In fact $95\%$ of sources have $E_8<0.117$. In considering the effect of extinction on colours it should be appreciated that the template colours themselves are not dereddened. Rather they are the median observed colours of stars in the BUD sample of \citet{schmidt15} within the UKIDSS footprint, {i.e.} the average colours of sources along a line of sight with $E_8\sim 0.035$. Hence it is appropriate to subtract this value of $E_8$ when computing the effects on classification caused by extinction. Therefore along a line of sight with $E_8=0.117$, the change in colour for any star in the sample is no greater than the change in colour produced by $E_8\sim 0.082$. This value of extinction corresponds to a reddening in the $i-J$ colour of 0.10 \citep{green18}. Since the change in $i-J$ between M7 and M9 is 0.93 \citep{skrzypek16}, the effect on classification is far less than half a spectral subclass. This means that for at least $95\%$ of the sample the effect of reddening on classification is negligible.

There are small areas of significant reddening, and 401 sources, or $1.2\%$ of the sample, have $E_8>0.2$. Stars in the sample along a line of sight with extinction of  $E_8$ are reddened in $i-J$ colour by up to 0.20 (following the argument above), equivalent to nearly half a spectral subclass. Therefore the classification for sources with $E_8>0.2$ should be treated as uncertain. These sources are denoted in the catalogue with a colon, i.e. M7:. The majority lie in two small regions within (but not filling) the bounds $56.0<\alpha<61.0,$ $-1.5<\delta<+1.5$  and $42.0<\alpha<57,$ $3.5<\delta<7.0$. Excluding these regions reduces the number of sources with $E_8>0.2$ from 401 to 89, i.e. removes $78\%$ of the reddened objects. The total sample size is reduced from 33665 to 32942, and the effective area of the survey is reduced from 3070 to 3031\,deg$^2$. Therefore by reducing the effective area by $1.3\%$, the proportion of significantly reddened objects in the sample is reduced from $1.2\%$ to $0.3\%$ of the total. We use the reduced sample in the analysis of the LF. We provide the survey solid angle, as a function of Galactic latitude $b$ in Table \ref{areas}, for the reduced effective area of 3031\,deg$^2$. 

\begin{table}\tiny
\centering
\caption{Quantiles of photometric errors and reddening}
\begin{tabular}{l c c c c c c c c c}
\hline\hline
 & $\sigma_{i}$ & $\sigma_{z}$ & $\sigma_{Y}$ & $\sigma_{J}$ &
$\sigma_{H}$ & $\sigma_{K}$ & $E_8$ \\  
 Median   & 0.033 & 0.038 & 0.020 & 0.016 & 0.020 & 0.024 & 0.035 \\
 $90\%$ & 0.060 & 0.062 & 0.032 & 0.028 & 0.035 & 0.040 & 0.088 
 \\ \hline 
\end{tabular}
\label{photomerrors}
\end{table}

\begin{table}\tiny
\centering
\caption{Solid angle of survey as a function of Galactic latitude. The area is zero at angles
not listed.}
\begin{tabular}{rrrrrrrrr}
\hline\hline
 $b_{min}$ & $b_{max}$ & area & $b_{min}$ & $b_{max}$ & area & $b_{min}$ & $b_{max}$ & area \\
                &                & deg$^2$ &                &                & deg$^2$ &                &                & deg$^2$ \\ \hline
   89 &  90 &  2.9 &  51 &  52 & 34.6 & -29 & -28 &  2.9 \\  
   88 &  89 &  8.2 &  50 &  51 & 35.7 & -30 & -29 &  2.9 \\
   87 &  88 & 11.3 &  49 &  50 & 37.6 & -31 & -30 &  2.7 \\
   86 &  87 & 14.3 &  48 &  49 & 38.4 & -32 & -31 &  2.8 \\
   85 &  86 & 16.0 &  47 &  48 & 38.1 & -33 & -32 &  2.9 \\
   84 &  85 & 16.5 &  46 &  47 & 40.2 & -34 & -33 &  3.0 \\
   83 &  84 & 15.1 &  45 &  46 & 42.3 & -35 & -34 &  2.9 \\
   82 &  83 & 16.8 &  44 &  45 & 42.2 & -36 & -35 &  3.1 \\
   81 &  82 & 16.3 &  43 &  44 & 42.5 & -37 & -36 &  3.1 \\
   80 &  81 & 13.8 &  42 &  43 & 44.6 & -38 & -37 &  2.9 \\
   79 &  80 & 13.6 &  41 &  42 & 40.5 & -39 & -38 &  3.4 \\
   78 &  79 & 16.1 &  40 &  41 & 32.6 & -40 & -39 &  6.5 \\
   77 &  78 & 22.4 &  39 &  40 & 29.5 & -41 & -40 & 12.6 \\
   76 &  77 & 27.1 &  38 &  39 & 26.4 & -42 & -41 & 20.1 \\
   75 &  76 & 30.4 &  37 &  38 & 22.7 & -43 & -42 & 25.9 \\
   74 &  75 & 33.1 &  36 &  37 & 19.5 & -44 & -43 & 29.6 \\
   73 &  74 & 31.2 &  35 &  36 & 18.7 & -45 & -44 & 35.7 \\
   72 &  73 & 27.9 &  34 &  35 & 19.6 & -46 & -45 & 45.2 \\
   71 &  72 & 30.3 &  33 &  34 & 19.3 & -47 & -46 & 58.2 \\
   70 &  71 & 32.9 &  32 &  33 & 19.8 & -48 & -47 & 63.3 \\
   69 &  70 & 37.0 &  31 &  32 & 19.1 & -49 & -48 & 59.7 \\
   68 &  69 & 38.1 &  30 &  31 & 19.9 & -50 & -49 & 56.5 \\
   67 &  68 & 41.8 &  29 &  30 & 20.3 & -51 & -50 & 52.1 \\
   66 &  67 & 44.3 &  28 &  29 & 20.6 & -52 & -51 & 50.0 \\
   65 &  66 & 45.1 &  27 &  28 & 21.1 & -53 & -52 & 45.1 \\
   64 &  65 & 47.5 &  26 &  27 & 22.2 & -54 & -53 & 41.0 \\
   63 &  64 & 49.4 &  25 &  26 & 21.4 & -55 & -54 & 43.8 \\
   62 &  63 & 52.7 &  24 &  25 & 22.2 & -56 & -55 & 44.2 \\
   61 &  62 & 56.9 &  23 &  24 & 22.2 & -57 & -56 & 40.1 \\
   60 &  61 & 59.5 &  22 &  23 & 17.2 & -58 & -57 & 37.1 \\
   59 &  60 & 57.9 &  21 &  22 & 11.2 & -59 & -58 & 33.9 \\
   58 &  59 & 53.5 &  20 &  21 &  6.3 & -60 & -59 & 30.0 \\
   57 &  58 & 46.5 &  19 &  20 &  2.1 & -61 & -60 & 24.1 \\
   56 &  57 & 44.7 & -24 & -23 &  1.3 & -62 & -61 & 12.1 \\
   55 &  56 & 41.8 & -25 & -24 &  2.8 & -63 & -62 & 13.2 \\
   54 &  55 & 41.8 & -26 & -25 &  2.9 & -64 & -63 &  5.2 \\
   53 &  54 & 39.1 & -27 & -26 &  2.9 &     &     &      \\
   52 &  53 & 36.0 & -28 & -27 &  2.9 &     &     &      \\ \hline 
\end{tabular}
\label{areas}
\end{table}

\begin{figure}
\centering
\includegraphics[width = 10.5cm, trim = 0.0cm 7cm 3.0cm 7.0cm]{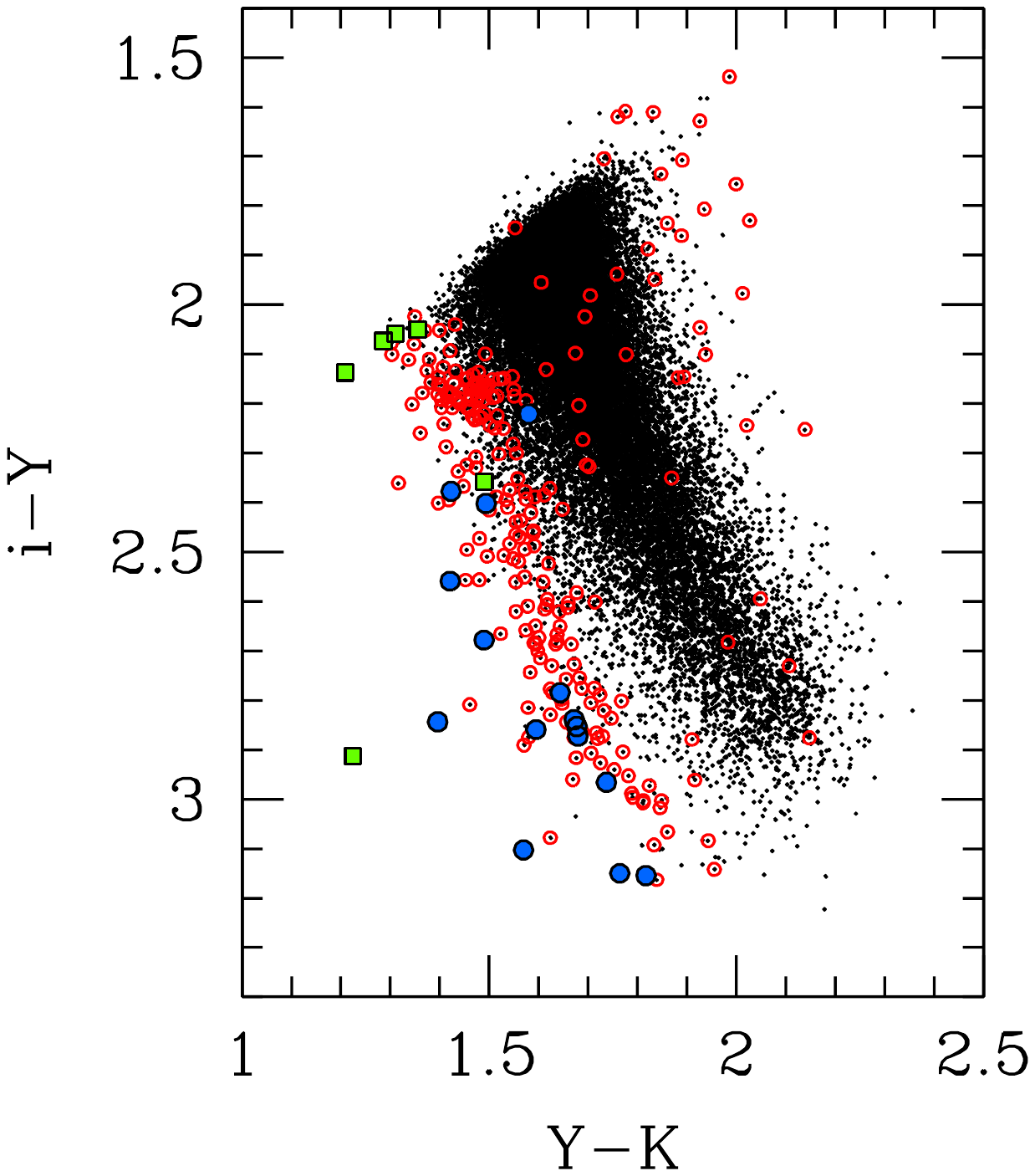}
\caption{Two-colour diagram $i-Y$ vs. $Y-K$ for the new sample. The
  large points are the known subdwarfs in the sample, listed in Table
  \ref{subdwarfs}: blue circles show types sdM and sdL, and green
  squares show type esdL. The red open circles indicate sources with
  $\chi^2>15$ and $E_8<0.2$.}
\label{twocol}
\end{figure}

\begin{figure}
\centering
\includegraphics[width = 10.5cm, trim = 1.0cm 9cm 0cm 7.0cm]{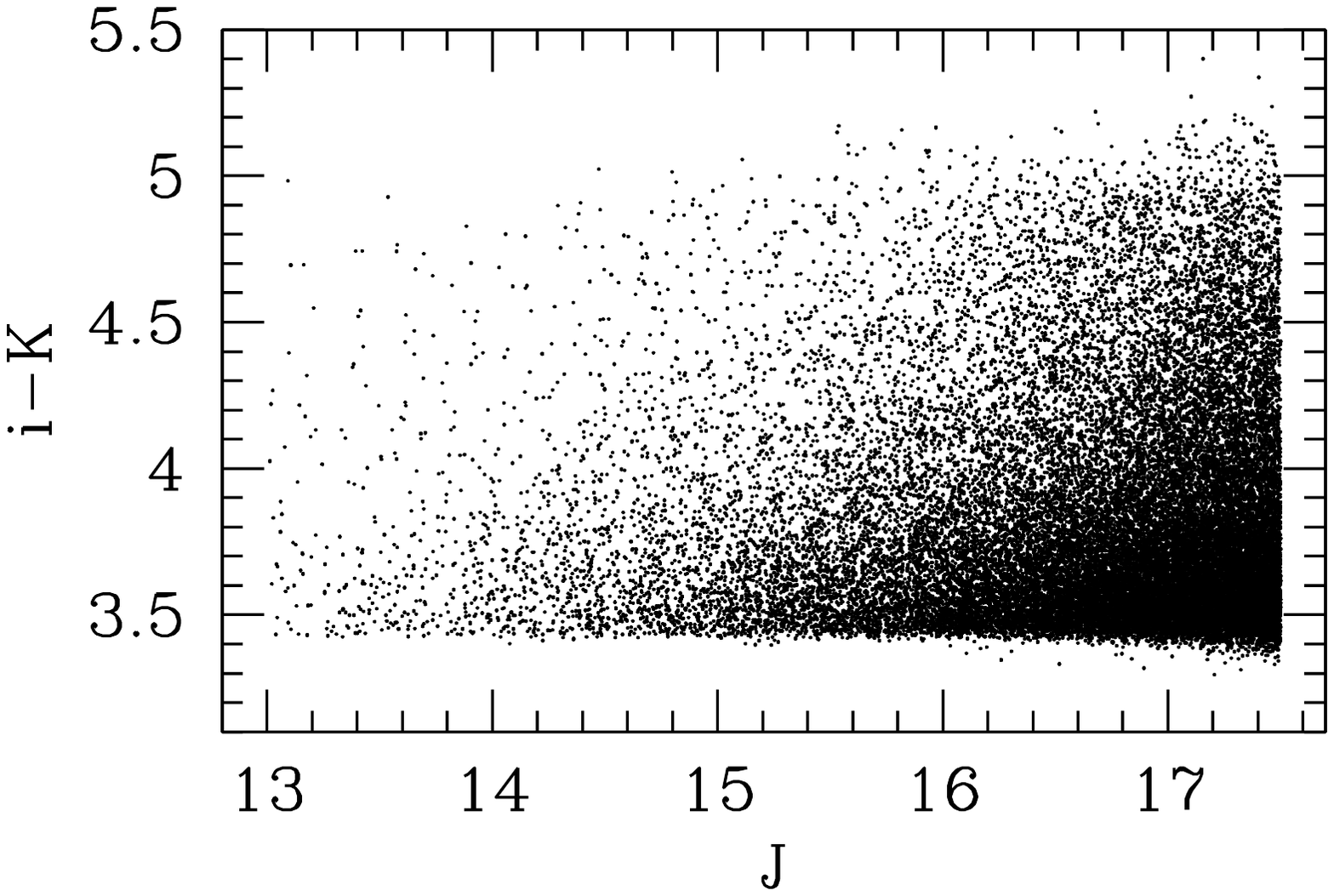}
\caption{Colour-magnitude diagram $i-K$ vs. $J$ for the new sample.}
\label{ikj}
\end{figure}

\begin{figure}
\centering
\includegraphics[width = 9.cm, trim = 1.5cm 10cm 0cm 7.0cm]{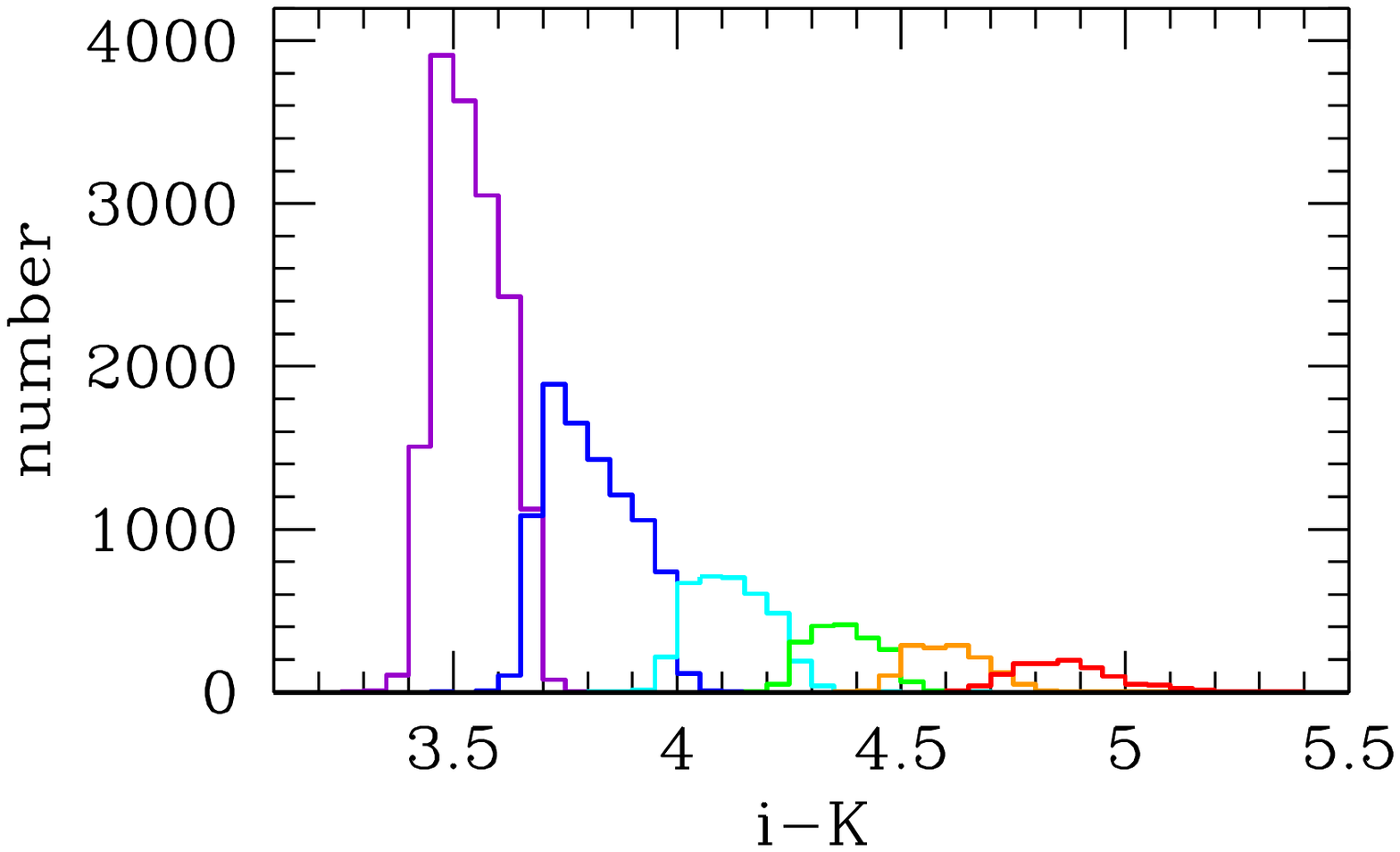}
\caption{Histograms of the distribution of $i-K$ colours for each subtype, in half subtype bins, from M7 (left, violet) to M9.5 (right,red).}
\label{ikhist}
\end{figure}

\begin{figure}
\centering
\includegraphics[width = 9.cm, trim = 1.5cm 10cm 0cm 8cm]{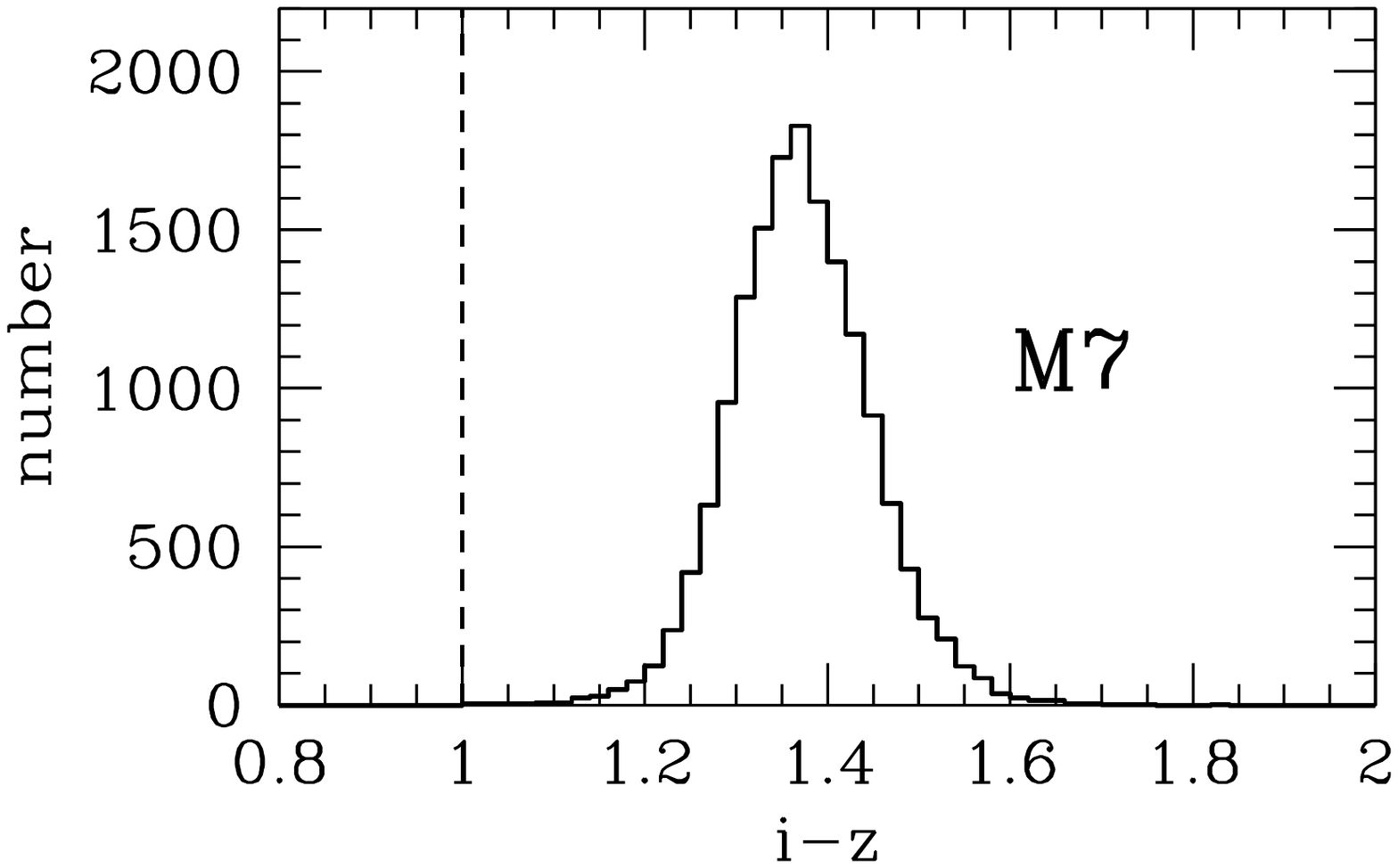}
\includegraphics[width = 9.cm, trim = 1.5cm 10cm 0cm 8cm]{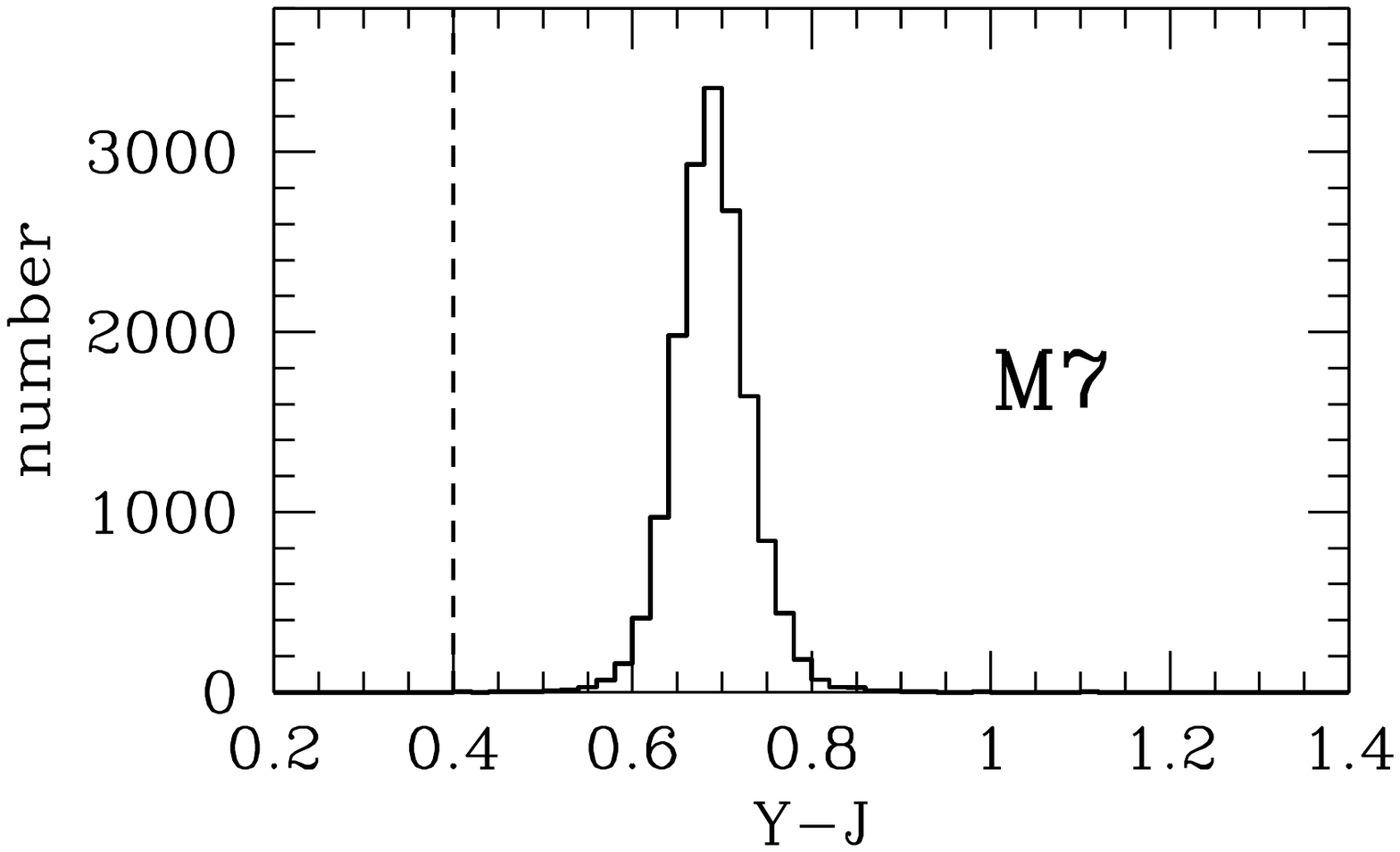}
\caption{Comparison of the colours of M7 dwarfs in the sample against the selection colour cuts.}
\label{izyjhist}
\end{figure}

For a small proportion of sources the SDSS uncertainties are larger than expected in comparison with other sources of similar brightness. These are objects for which the SDSS deblending algorithm has boosted the uncertainties relative to the Poisson values. We identified 582 sources, or $1.7\%$ of the total, where either the $i$ band uncertainty was $>0.20$ or the $z$ band uncertainty was $>0.15$ (or both). For these sources the classification is denoted as uncertain. The majority of cases are stars with a nearby brighter companion, causing difficulty in deblending. Many of the objects could be members of close binary systems. Because of the large uncertainties the bands with bad photometry effectively do not contribute to the classification, which should otherwise be reliable.  Because the photometry in the affected band is not useful, we removed these sources from the plots presented in Figs \ref{twocol} to \ref{izyjhist}. 

In this spectral range M7-M9.5, for each colour $i-z$, $z-Y$, $Y-J$, $J-H$, $H-K$, there is an approximately linear relation between colour and spectral type; the later types are redder. For this reason the stellar sequence presents a linear relation in a two-colour diagram. This is illustrated in Fig. \ref{twocol}, which plots $i-Y$ versus $Y-K$. Therefore the $i-K$ colour is the single colour most closely correlated with spectral type. In Fig. \ref{ikj} we plot $i-K$ versus $J$ for the new sample. This figure illustrates the steep increase in number towards fainter magnitudes and the steep decrease in number counts towards redder colours. The colour $i-K$ on its own provides a good measure of spectral type, as illustrated in Fig. \ref{ikhist}, where the histograms of the $i-K$ colour range of each spectral type are presented, showing little overlap between spectral types.

In the sample selection (\S\ref{selection}), colour cuts $i-z>1.0$, $Y-J>0.4$ were applied before classifying. In Fig. \ref{izyjhist} we plot histograms of these two colours for the earliest spectral classification in the sample, M7, i.e. plotting the bluest objects in the new sample. It is clear from both histograms that the number of sources lost from applying these colour cuts is negligible.

\begin{figure}
\centering
\includegraphics[width = 10.cm, trim = 0.5cm 7cm 0cm 3cm]{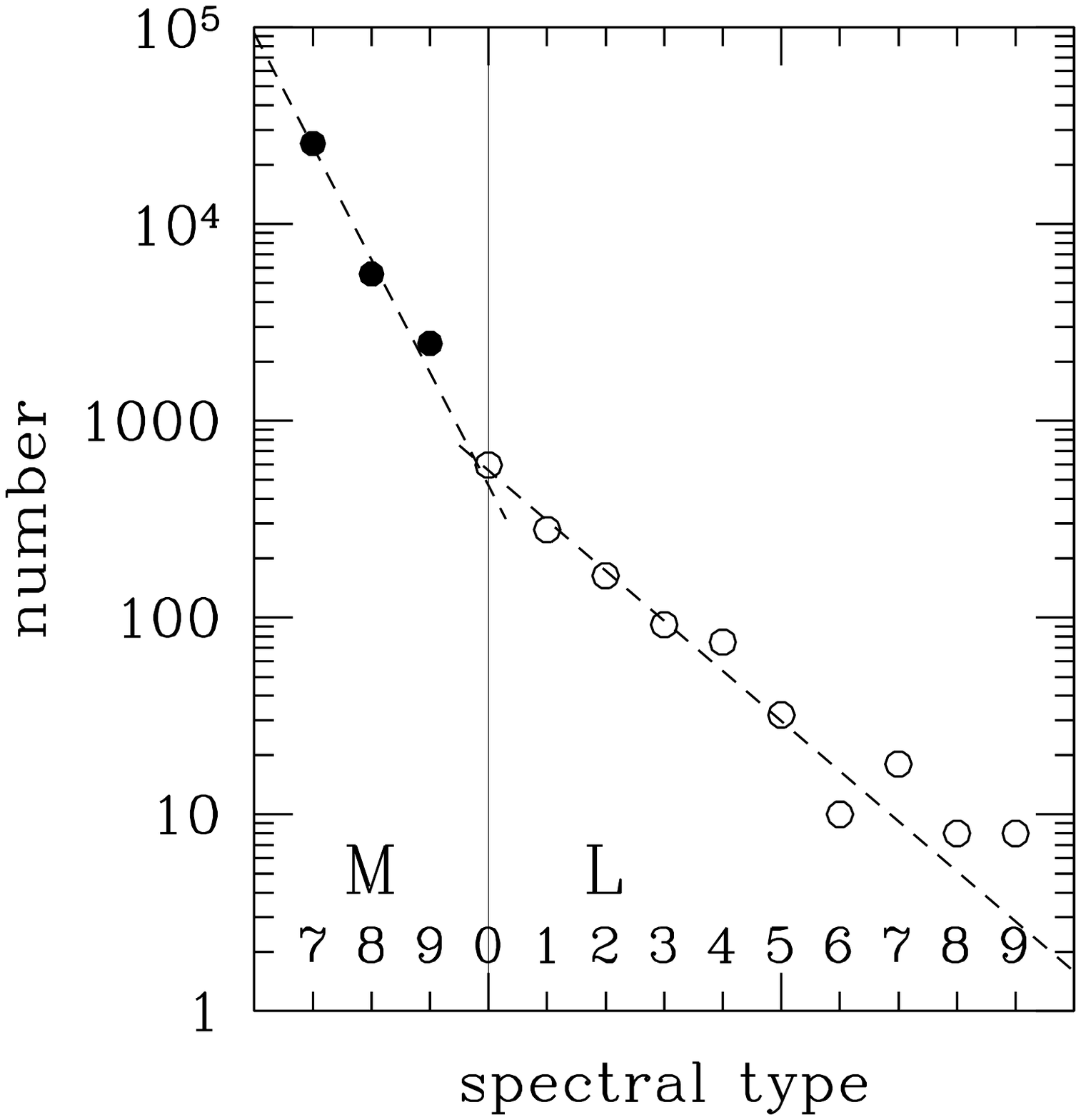}
\caption{Number counts of M7-9 dwarfs (solid symbols) compared to number counts for L dwarfs (open symbols) over the same area and depths. The dashed lines indicate the log-linear fits referred to in the text.}
\label{countsfit}
\end{figure}

The numbers of sources broken down into half subtype bins are listed in Table \ref{counts}.
 The numbers of sources per full spectral bin (e.g. M7 and M7.5 are combined as M7) are plotted against spectral type and compared against the counts of L dwarfs from \citet{skrzypek16}, which are for the same area and magnitude cuts. The numbers decline towards later spectral types. The decline is steeper for the late M dwarfs and less steep for the L dwarfs. There is a break in the slope that occurs at L0. We assume a functional form $N=10^{a-bs}$, i.e. the counts are linear in log$_{10}$. In this case $s$ is spectral type, numbering M7 as 7 to L9 as 19, and $a$ and $b$ are constants.  We fit
       to the counts separately for M7 to L0 and for L0 to L9, finding $b_M=0.57$ for the range M7 to L0, and $b_L=0.25$ for the range L0 to L9. These fits are plotted in Fig.\ref{countsfit}.
  
\begin{table}
\centering
\caption{Number counts by spectral type}
\begin{tabular}{l r}
\hline\hline
{SpT} & {Count}  \\ \hline
M7    & 16202 \\
M7.5 & 9436 \\
M8    & 3680 \\
M8.5 & 1877 \\
M9    & 1369 \\
M9.5 & 1101 \\ \hline
\end{tabular}
\label{counts}
\end{table}

\begin{figure}
\centering
\includegraphics[width = 9.cm, trim = 1.5cm 10cm 0cm 8cm]{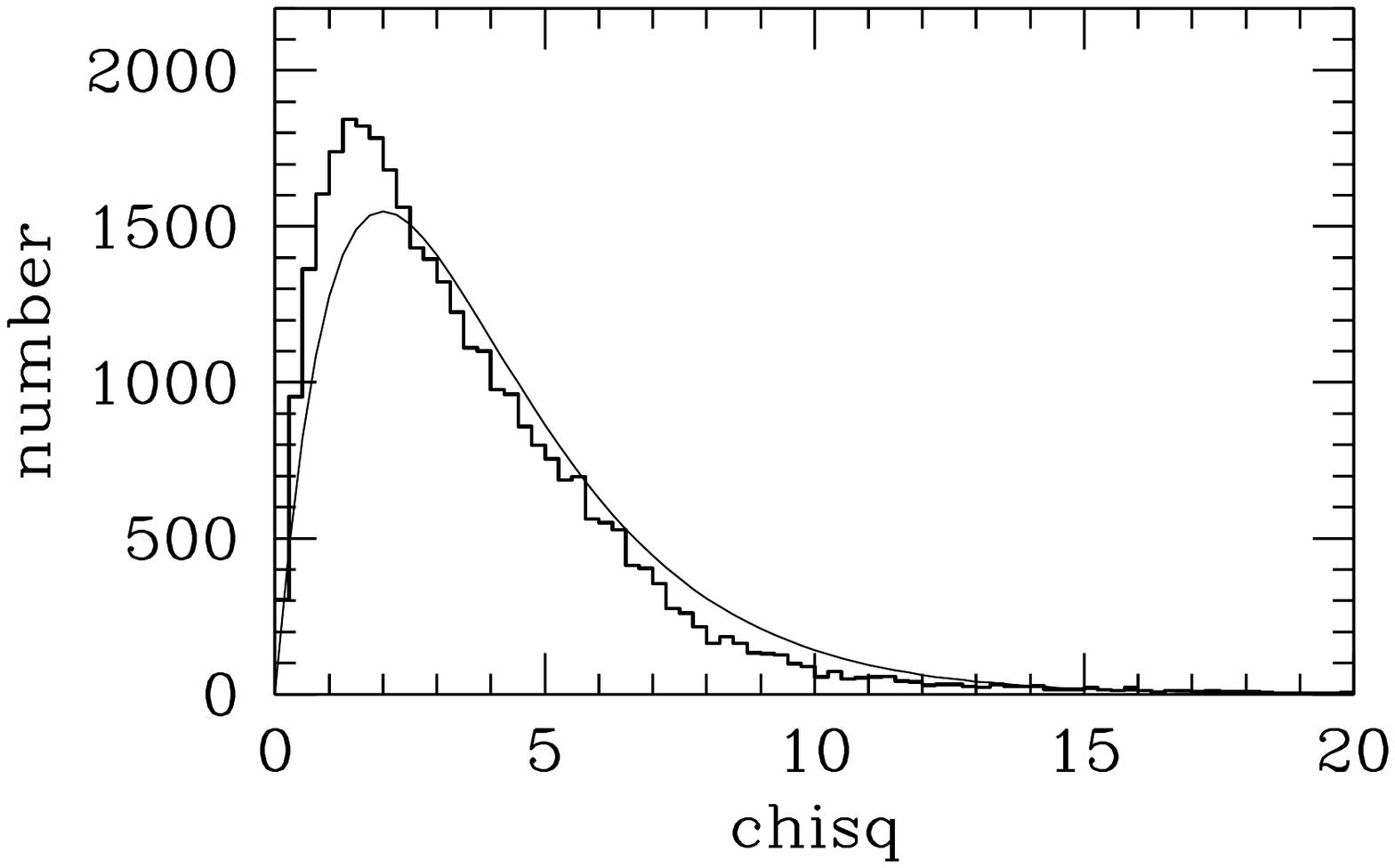}
\caption{Distribution of $\chi^2$ for the full sample compared to the theoretical distribution for 4 degrees of freedom.}
\label{chihist}
\end{figure}

 There are a number of sources with large values of $\chi^2$, because
 the colours are a poor match to the templates. These sources are
 potentially interesting. The $\chi^2$ distribution for the sample is
 plotted in Fig. \ref{chihist}. Each SED has six data points and there are two parameters to fit: the
 brightness (i.e. overall normalisation) and the spectral type,
 leaving four degrees of freedom. Overplotted on the $\chi^2$
 histogram is the theoretical $\chi^2$ distribution for four degrees
 of freedom. The curve is a reasonable fit, but the data are skewed to
 slightly smaller values, indicating that the uncertainties are
 slightly overestimated. The reason for the poor fit probably lies in
 the way the spread in the properties of the population has been
 modelled. In fitting templates to each SED, an additional uncertainty
 of 0.05 mag is added in quadrature to the photometric error in each
 band, to model the spread in each colour of the population, at each
 spectral type. This is a simplification, since it treats each band as
 independent, whereas the actual spread in colours is
 characterised by correlations between bands. The theoretical $\chi^2$
 curve is nevertheless useful for comparison. If the errors were
 modelled perfectly, and the source list contained no peculiar objects,
 we would expect 158 (17) sources with $\chi^2>15$ (20),
 respectively, whereas the actual numbers are 298 (108) or 260 (87) if we exclude
 sources with $E_8>0.2$, which are presumably peculiar because of the
 large reddening. With this in mind we selected $\chi^2>20$ as
 indicating that a source is peculiar, and the classifications are
 marked e.g. M7p. We scrutinised the images of these 108
 sources in all bands, and it appears that all the sources are
 genuinely peculiar. Nevertheless it is in the nature of the selection
 to pick out sources with incorrect photometry, so there could be some
 remaining errors.
  
 Several of the 87 peculiar (and not reddened) sources are known
 subdwarfs, including the two sources with the largest values of
 $\chi^2=111,62$. To investigate this further we matched the large
 sample of L subdwarfs in Table 1 of \citet{zhang18}, as well as
 additional M subwdarfs in their Table 6, to our full sample. The 19
 matched L subdwarfs and 2 matched M subdwarfs are listed in Table
 \ref{subdwarfs}. Six of these are classified esdL, which are of lower
 metallicity than the sdL class \citep[for details of the
   classification scheme, see][]{zhang17}.  On average the
   photo-type classifications are two subtypes earlier than the
 correct spectral classification because of their unusual blue
 colours. Their colours are plotted in Fig. \ref{twocol}. The $\chi^2$
 values for this sample range from 9 to 111 and have an average value of
 32. Of the 21 matched subdwarfs, 15 have $\chi^2>20$, or $17\%$ of
 the 87 peculiar sources that are not highly reddened. It would
 therefore be interesting to investigate further the remaining 72
 sources. They might include subdwarfs missed by proper motion
 selection.

We can use the subdwarfs as a guide to estimate how many objects are significantly misclassified because of their peculiar colours.
Of the 21 known subdwarfs, 19, that is nearly all, have $\chi^2>15$. The other two subdwarfs, with lower $\chi^2$, are misclassified by only 0.5 and 1 subtypes. 
There are only 260 sources in total with $\chi^2>15$, $E_8<0.2$. These are plotted as red circles in Fig. \ref{twocol}. The majority, like the subdwarfs,  lie to the left of the sequence visible in Fig. \ref{twocol}, but have on average less extreme colours. The selection $\chi^2>15$ therefore probably identifies nearly all sources misclassified by two subtypes or more, as well as intermediate objects. This suggests that the proportion of blue objects misclassified by two subtypes or more is safely less than $1\%$. The proportion of red circles lying to the right of the sequence in Fig. \ref{twocol} is considerably smaller. These may represent dwarfs of spectral type earlier than M7 that are selected as M7 or later because they are peculiar and red.

\begin{table}\tiny
\centering
\caption{Catalogued subdwarfs in the M dwarf sample}
\begin{tabular}{lrll}
\hline\hline
  \multicolumn{1}{c}{name}   & \multicolumn{1}{c}{$\chi^2$}  & PhT & SpT \\ \hline
ULAS J002009.36+160451.2 &  19.78 & M7.5  & sdM9     \\
ULAS J021258.08+064115.9 &  25.16 & M8p   & sdL1     \\
ULAS J023803.12+054526.2 &  27.73 & M8p   & sdL0     \\
ULAS J033351.11+001405.9 &  30.74 & M7p   & esdL0     \\
ULAS J082206.61+044101.9 &  36.28 & M8.5p & sdL0     \\
ULAS J124425.76+102439.3 &  60.84 & M7p   & esdL0.5     \\
ULAS J124947.05+095019.9 &  26.71 & M7.5p & sdL1     \\
ULAS J125226.63+092920.1 &  18.68 & M8.5  & sdL0     \\
ULAS J133348.27+273505.6 &  37.25 & M8p   & sdL1     \\
ULAS J134206.87+053725.0 &  20.21 & M9p   & sdL0.5     \\
ULAS J134749.80+333601.7 &  62.42 & M8.5p & sdL0     \\
ULAS J134852.93+101611.9 &  23.86 & M9p   & sdL0     \\
ULAS J135359.58+011856.8 &  13.87 & M9    & sdL0     \\
ULAS J141405.67-014204.1 &  24.18 & M7p   & esdL0     \\
ULAS J141832.36+025323.1 &  25.24 & M9p   & sdL0     \\
ULAS J143517.19-014713.2 &   8.96 & M7.5  & sdM8     \\
ULAS J145234.66+043738.5 &  16.24 & M8    & esdL0.5     \\
ULAS J151913.04-000030.1 & 111.97 & M8p   & esdL4     \\
ULAS J225902.15+115602.1 &  19.75 & M9    & sdL0     \\
ULAS J230256.54+121310.3 &  26.74 & M8p   & sdL0     \\
ULAS J231924.36+052524.6 &  38.25 & M7p   & esdL1     \\
\end{tabular}
\label{subdwarfs}
\end{table}

The sample is limited to objects classified as point sources in
UKIDSS. Point sources include unresolved binaries. A significant
proportion of the sources appear in the {\em Gaia} DR2 catalogue and
the parallaxes, combined with the apparent magnitudes in our
catalogue, are useful to identify which sources are unresolved
binaries. Any such objects in our sample with small values of $\chi^2$
are likely to be pairs of dwarfs of similar spectral type. Unresolved
binaries with large $\chi^2$ may comprise a M dwarf primary and a
secondary of later spectral type so that the colours are dominated by
the primary. Another possibility is a M dwarf with a cool white dwarf
(WD) companion. A large sample of M+WD binaries has been compiled by
\citet{rebassa13} also using SDSS+UKIDSS. The majority of the M stars
in their sample are M2 and M3. Therefore any white dwarfs in M+WD
binaries found in our new sample might have different characteristics
to the white dwarfs in their sample. The source ULAS
J115908.00+103944.0, which is the object with the fifth largest value
of $\chi^2$ in our sample, could be one example. The source was
selected by SDSS for spectroscopic observation as a quasi-stellar object candidate, but
the spectrum is classified M7 and displays excess blue continuum
light. The colours of the source are such that it does not satisfy the
selection criteria of \citet{rebassa13}.

\section{Precision of spectral types}
\label{accuracy}

We can establish the precision of the photo-type classifications by comparing them against classifications in the BUD sample \citep{schmidt15} that was used to establish the relations between colour and spectral type in this range \citep{skrzypek16}. 
The precision of the classification impacts the number counts due to Eddington bias \citep{eddington}, i.e. the effect that because of the steep relation between counts and spectral type the uncertainty in classification scatters more objects from earlier to later classifications than vice versa. By establishing the precision it is then possible to correct the number counts for this effect.

\begin{figure}
\centering
\includegraphics[width = 9.cm, trim = 2cm 8cm 0cm 5.5cm]{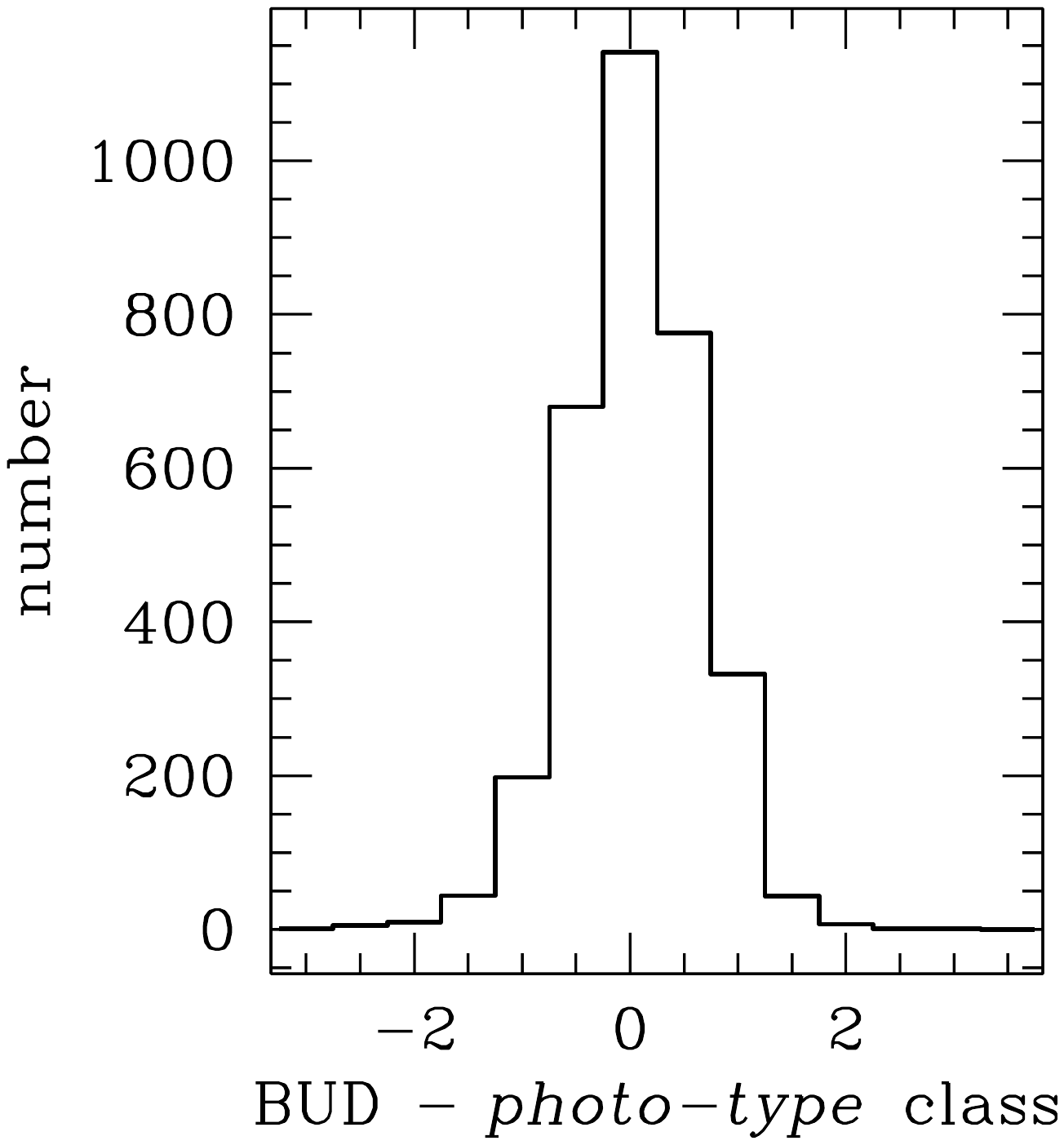}
\caption{Histogram of the difference in the classification between spectroscopy and using photo-type for the 3\,239 M7 to M9 dwarfs from \citet{schmidt15} that matched the classified parent sample.}
\label{schmidthist}
\end{figure}

The BUD classifications are measured to the nearest full subtype. The sample includes spectra of 11820 M7 to L8 dwarfs. We matched the M7 to M9 dwarfs in this sample to the classified parent sample of 404\,496 sources (\S\ref{selection}) from which our sample of M dwarfs is drawn, finding 3239 matches within the UKIDSS footprint and brighter than $J=17.5$. To establish the precision it is important to match to the parent sample since some of the BUD M7 dwarfs are classified by photo-type as earlier than M7. We then measure the mean and scatter of the difference in classification, BUD minus photo-type. The histogram of differences is plotted in Fig. \ref{schmidthist}. The mean value for the 3239 dwarfs is 0.05 subtypes. Since we assume that the BUD classifications are in the mean correct, this establishes that the photo-type classifications are extremely accurate (systematic error). This is not surprising since, of course, the BUD sample itself was used to measure the template colours. The scatter in the differences establishes the precision (random error). The standard deviation of the differences in classification is just 0.6 subtypes. The same value is obtained whether the scatter is measured about the mean, or about zero, and whether the handful of sources with large differences in classification (greater than 2 subtypes) is clipped, or not. 

This scatter is remarkably small considering that it is made up of
three contributions added in quadrature: 1) the precision of photo-type; 2) the precision of the spectroscopic classifications; and
3) a contribution of $0.5/\sqrt{3}=0.3$ solely from the quantisation
of the spectroscopic classifications into whole subtypes, rather than
half subtypes. This means that both the photo-type and the
spectroscopic classifications have a precision of better than 0.5
subtypes {rms}. This precision is as good as the precision of the
best automated spectral classifiers \citep[e.g.][]{Christlieb02}. For
a power-law slope of the number counts of 0.57 (Fig. \ref{countsfit}),
and a precision of 0.5 subtypes, the Eddington bias is at the level
of $20\%$, i.e. the number counts are too large by a factor
1.2. Nevertheless it is debatable whether this calculation of
Eddington bias has much meaning. If the sample had been obtained by a
spectroscopic campaign with classifications to this precision it is
unlikely any correction to the number counts would be deemed
necessary, and therefore we ignore this correction in computing
the LF.

\section{Summary}
\label{summary}

In this paper we have presented a homogeneous sample of 33\,665 bright
$J<17.5$ M7 to M9.5 dwarfs, which have accurate spectral types obtained by
applying the photo-type method to $izYJHK$ SDSS and UKIDSS
photometry. The effective area of the survey is 3070\,deg$^2$. The
sample is of high S/N and low reddening. We took care to
include, where possible, dwarfs that are located at small angular
separations from bright stars. These may be binary companions to the
bright stars, and are sometimes excluded in working with SDSS
data. The sample is a companion to the sample of 1361 L and T dwarfs
provided by \citet{skrzypek16}, selected from the same multicolour
dataset, and to the same depth. The number counts as a function of
spectral type fall steeply over the range M7 to M9.5 towards later
types and there is a break at L0 to a flatter relation in the L
dwarfs. For each source we list coordinates, $izYJHK$ photometry,
Galactic reddening, and the $\chi^2$ of the six-band photometric fit, in
addition to the photo-type classification. The classifications
are provided to the nearest half subtype and are precise to better
than 0.5 subtypes rms. We argued that the precision is so
good that Eddington bias in the number counts as a function of
spectral type may be disregarded. The sources with large $\chi^2$
include subdwarfs, probably dwarfs of intermediate metallicity,
and other peculiar types. 

All the sources lie within a distance of 235\,pc, so the sample will
be useful for measuring the structure of the Milky Way disc close to
the Galactic plane, and it will provide a new, more accurate,
measurement of the space density of M7 to M9.5 dwarfs. To
measure the local space density we must measure the variation of the
space density with height $\left| z \right|$ from the Galactic plane
and extrapolate to $z=0$. The variation of the density of stars with
height from the plane is often characterised using a
$\mathrm{sech}^2\left(\frac{\modz}{2z_s}\right)$ function
\citep{spitzer42}. The function is exponential $e^{-\modz/z_s}$ at
large heights but softens such that the central-plane density is
reduced by a factor of four compared to the extrapolation to
the central plane of the exponential distribution. While the
$\mathrm{sech}^2$ function has been widely used, there is very little
observational evidence of the actual softening near the Galactic
plane. The new sample will be very useful to examine the form of the density
distribution at small heights from the Galactic plane.

\begin{acknowledgements}
This work was supported by Grant ST/N000838/1 from the Science and Technology Facilities Council.
The UKIDSS project is defined in \citet{UKIDSS}. UKIDSS used the UKIRT Wide Field
Camera \citep{casali07}. The photometric system is described in
\citet{hewett06}, and the calibration is described in
\citet{hodgkin09}. The science archive is described in
\citet{hambly08}.\\
Funding for SDSS-III has been provided by the Alfred P. Sloan
Foundation, the Participating Institutions, the National Science
Foundation, and the U.S. Department of Energy Office of Science. The
SDSS-III web site is http://www.sdss3.org/.  SDSS-III is managed by
the Astrophysical Research Consortium for the Participating
Institutions of the SDSS-III Collaboration including the University of
Arizona, the Brazilian Participation Group, Brookhaven National
Laboratory, Carnegie Mellon University, University of Florida, the
French Participation Group, the German Participation Group, Harvard
University, the Instituto de Astrofisica de Canarias, the Michigan
State/Notre Dame/JINA Participation Group, Johns Hopkins University,
Lawrence Berkeley National Laboratory, Max Planck Institute for
Astrophysics, Max Planck Institute for Extraterrestrial Physics, New
Mexico State University, New York University, Ohio State University,
Pennsylvania State University, University of Portsmouth, Princeton
University, the Spanish Participation Group, University of Tokyo,
University of Utah, Vanderbilt University, University of Virginia,
University of Washington, and Yale University.
\end{acknowledgements}
\bibliographystyle{aa}
\bibliography{samplev1}

\end{document}